%% file: hepNewBC.tex
\documentclass[12pt,a4paper]{article}
\usepackage{amsmath,amsfonts,epsfig}

\newcommand{\Real}{\mathbb{R}}
\newcommand{\Integer}{\mathbb{Z}}
\newcommand{\Natural}{\mathbb{N}}
\newcommand{\eq}{\begin{eqnarray*}}
\newcommand{\qe}{\end{eqnarray*}}
\newcommand{\eqn}{\begin{eqnarray}}
\newcommand{\qen}{\end{eqnarray}}
\newcommand{\mat}{\begin{pmatrix}}
\newcommand{\tam}{\end{pmatrix}}
\newcommand{\tr}{\mbox{tr}}
\def\a{\alpha} 

\newcommand{\mf}{\mathfrak} 
\newcommand{\mc}{\mathcal} 

\newcommand{\rrangle}{\rangle\!\rangle}
\newcommand{\llangle}{\langle\!\langle}
\newcommand{\id}{\text{\bf id}}
\newcommand{\p}{\mf{p}}
\newcommand{\g}{\mf{g}}
\newcommand{\h}{\mf{h}}
\newcommand{\ap}{\hat{\mf{p}}}
\newcommand{\ag}{\hat{\mf{g}}}
\newcommand{\ah}{\hat{\mf{h}}}
\newcommand{\au}{\widehat{\mf{u}}}
\newcommand{\asu}{\widehat{\mf{su}}}

\newcommand{\A}{\mc{A}}
\newcommand{\G}{{\text{G}}}
\renewcommand{\H}{{\text{H}}}
\newcommand{\embin}{\hookrightarrow}
\renewcommand{\P}{{\text{P}}}

\newcommand{\SU}{\text{SU}}
\newcommand{\U}{\text{U}}
\newcommand{\B}{\mc{B}}
\newcommand{\RA}{\text{Rep}(\mc{A})}
\newcommand{\Rep}{\text{Rep}}  
\newcommand{\Spec}{\text{Spec}}
\newcommand{\Hilb}{\mc{H}}
\newcommand{\Hilbb}{\bar{\mc{H}}}
\newcommand{\Gid}{\mc{G}_{\text{id}}}
\newcommand{\Allowed}{\text{All}}

\newcommand{\Cent}{\mc{Z}}

\title{Symmetry Breaking Boundary States \\[1mm] and Defect Lines}

\author{Thomas Quella\footnote{{\tt email: quella@aei.mpg.de}} 
   \ and \ Volker 
   Schomerus\footnote{{\tt email: vschomer@aei.mpg.de}}\\[2mm] 
MPI f\"ur Gravitationsphysik, Albert-Einstein-Institut\\ 
Am M\"uhlenberg 1, D-14476 Golm, Germany}

\date{March 26, 2002}
  
\begin{document}
\maketitle
\baselineskip16pt 

\begin{abstract}
We present a large and universal class of new boundary states 
which break part of the chiral symmetry in the underlying bulk 
theory. Our formulas are based on coset constructions and they 
can be regarded as a non-abelian generalization of the ideas
that were used by Maldacena, Moore and Seiberg to build new 
boundary states for~$\SU(N)$. We apply our expressions to construct
defect lines joining two conformal field theories with possibly 
different central charge. Such defects can occur e.g.\ in 
the AdS/CFT correspondence when branes extend to the boundary 
of the AdS-space. 
\end{abstract}

\vspace*{-15cm}\noindent
{AEI-2002-024 \hfill hep-th/0203161}\\
{PAR-LPTHE-02-09}

\newpage

\section{Introduction}

During the last years, the microscopic techniques of boundary 
conformal field theory have been developed into a powerful 
tool that allows to study D-branes in curved backgrounds with 
finite curvature. Most of these investigations, however, focus 
on boundary conditions that preserve the full chiral symmetry in 
the bulk theory. The latter is typically much larger than the 
Virasoro algebra that must be unbroken to guarantee conformal 
invariance. Constructing boundary theories with the minimal 
Virasoro symmetry tends to lead into non-rational models which 
are notoriously difficult to control. 
\smallskip

Nevertheless, some progress has been made in this direction. 
Boundary conditions with the minimal Virasoro symmetry were 
systematically investigated for 1-dimensional flat targets
\cite{Gaberdiel:2001xm,Gaberdiel:2001zq,Janik:2001hb,
Cappelli:2002wq}. In-spite 
of this remarkable progress, such a complete control over conformal 
boundary conditions should be considered exceptional and it 
is probably very difficult to achieve for more complicated 
backgrounds. Less ambitious programs focus on intermediate 
symmetries which are carefully selected so as to render the 
boundary theory rational. 
\smallskip 

One possibility is to work with orbifold chiral algebras. This 
has been explored in great detail by several groups (see e.g.\ 
\cite{Fuchs:1999zi,Birke:1999ik} and also \cite{Behrend:1999bn})  
and it has lead to new boundary theories in group manifolds and 
other backgrounds. More recently, Maldacena, Moore and Seiberg 
\cite{Maldacena:2001ky,Maldacena:2001xj} have proposed further 
symmetry breaking boundary states for the $\SU(N)$ WZW-model. 
Their construction employs the chiral algebra of the $\SU(N)/
\U(1)^{N-1}$ coset theory (see also \cite{Rajaraman:2001ew}
for a similar analysis in a non-compact background). Our aim here 
is to turn these ideas into a more general procedure that provides
a large class of new boundary theories. The construction involves 
coset chiral algebras with non-abelian denominators and after some 
more technical refinements it has a good chance even to exhaust all 
rational boundary theories.
\medskip

To make this paper self-contained, we shall start our exposition 
with some background material on 2D boundary conformal field theory 
and on coset chiral algebras. This then enters crucially into 
our construction of the new boundary theories in the third 
section. After presenting formulas for the boundary states and 
computing the associated open string spectra we discuss the 
relation with the D-branes constructed in \cite{Maldacena:2001ky,
Maldacena:2001xj}. Finally, we shall sketch how our new states 
can be used to describe defect lines separating two different 
conformal field theories. Such systems have been studied by various 
authors (see e.g.\ \cite{McAvity:1995zd,Oshikawa:1997dj,
LeClair:1997gz,Nayak,Saleur:1998hq,Saleur:2000gp,Erdmenger:2002ex}) 
and they are known to appear e.g.\ in the context of the 
$AdS_3/CFT_2$ correspondence where $AdS_2$-branes can end 
on the boundary of $AdS_3$ \cite{Bachas:2001vj} (see also 
\cite{Karch:2000ct,Karch:2000gx,DeWolfe:2001pq}). In fact, 
it was mainly the interest in the latter that has motivated 
the present work even though the results we present can 
have many other applications.

\section{Background from Conformal Field Theory} 

In this section we collect some background material about 2D boundary 
conformal field theory (BCFT) and coset chiral algebras. One of the 
main aims is to set up the notations we are using throughout this work. 
Readers who are familiar with the relevant techniques from conformal 
field theory may skip this section and consult it only to look up 
our conventions.  

\subsection{Some boundary conformal field theory}

Let us start by reviewing some basic elements of (boundary) conformal 
field theory. Our presentation will closely follow the 
reference~\cite{Behrend:1999bn}. The central ingredient in any CFT 
is its chiral algebra $\A$ which contains the Virasoro algebra. 
We shall restrict ourselves to the so-called rational 
algebras $\A$ possessing a finite set $\RA$ of `physical' irreducible
representations. Furthermore, we assume that the two chiral 
algebras $\A$ and $\bar{\A}\cong\A$ of the bulk theory are identical.   
This is not the most general situation as there exist also so-called 
heterotic CFTs with different left and right-moving chiral algebras  
(see~\cite{Gannon:1993zq,Gannon:1994vp} for instance).%
\smallskip%

To fully specify the bulk CFT we still need to characterize its field
content. The space of fields decomposes into irreducible representations
for the product of the two chiral algebras,  
\eq
  \Hilb \ =\ \bigoplus_{\mu,\bar{\mu}\in\RA}\ Z^{\mu\bar{\mu}}
               \  \Hilb_\mu \, \otimes\, \bar{\Hilb}_{\bar{\mu}}
\qe
with some numbers $Z^{\mu\bar{\mu}}\in\Natural_0$. We call the set of 
all pairs $(\mu,\bar{\mu})$ that contribute to $\Hilb$ (including the 
multiplicities) the {\em spectrum} of the (bulk) theory and denote it 
by
\eq
  \Spec\:=\:
  \bigl\{\:(\mu,\bar{\mu}|\eta)\:\bigl|\:\eta=1,
  \ldots,Z^{\mu\bar{\mu}}\:\bigr\}\ \ .
\qe
Since time translation in the usual radial quantization is generated 
by the sum $L_0 + \bar{L}_0$ of the zero modes of the chiral Virasoro
fields, the spectrum of the theory may be captured by the toroidal 
partition function
\eq
  Z(q,\bar{q})
  \ =\ \tr_{\Hilb}\, q^{L_0-c/24}\, \bar{q}^{\bar{L}_0-c/24}
  \ = \sum_{\mu,\bar{\mu}\in\RA}\, 
  Z^{\mu\bar{\mu}}\ \chi_\mu(q)\,\bar{\chi}_{\bar{\mu}}(\bar{q})
\qe
where the argument $q=\exp(2\pi i\tau)$ is determined by the modulus 
$\text{Im }\tau>0$ of the torus. The number $c$ is the central charge 
of the Virasoro algebra and the characters of the chiral algebra are 
defined by
\eq
  \chi_\mu(q)\ =\ \tr_{\Hilb_\mu}\, q^{L_0-c/24}\ \ .
\qe
Consistency requires the partition function to be invariant under
modular transformations $T:\tau\mapsto\tau+1$ and $S:\tau\mapsto-1/
\tau$ which may be represented unitarily on the characters as 
follows
\eq
  T\chi_\mu(\tau)&=&\chi_\mu(\tau+1)\:=
  \:e^{2\pi i(h_\mu-c/24)}\,\chi_\mu(\tau)\\[2mm]
  S\chi_\mu(\tau)&=&\chi_\mu(-1/\tau)\:=\:
  \sum_{\nu\in\RA}S_{\mu\nu}\,\chi_\nu(\tau)
\qe
where we introduced the conformal weight $h_\mu$. For later use it is 
convenient to summarize some properties of the modular S-matrix, 
\eqn
  \label{eq:SMatrixProperties}
  S_{\mu\nu}\ =\ S_{\nu\mu}
  \qquad\qquad 
  S_{\mu^+\nu}\ =\ \bar{S}_{\mu\nu}
  \qquad\qquad
  \sum_{\lambda\in\RA}\bar{S}_{\mu\lambda}\,S_{\nu\lambda}
  \ =\ \delta_\nu^\mu\ \ .
\qen
Imposing invariance of the spectrum under modular transformations gives 
severe restrictions on the numbers $Z^{\mu\bar{\mu}}$. Nevertheless, 
there exist choices $Z^{\mu\bar{\mu}}=\delta^{\mu\bar{\mu}}$ and 
$Z^{\mu\bar{\mu}}=\delta^{\mu\bar{\mu}^+}$, the so-called diagonal
and charge conjugation invariants, which are always allowed.
\medskip 

We are interested in boundary conditions which preserve the chiral 
algebra~$\A$. When we specify boundary theories through the associated 
boundary states $|B\rangle$, the choice of the boundary condition 
is implemented by gluing conditions of the form 
\eqn
  \label{eq:BoundaryCondition}
  \bigl(\phi(z)-\Omega\bar{\phi}(\bar{z})\bigr)|B\rangle=0
  \qquad\text{ for }\qquad z=\bar{z}\ \ .
\qen
Here, the reflection of left into right movers is described by a 
gluing automorphism $\Omega\in\text{Aut}(\A)$ which must leave the
energy momentum tensor invariant in order to preserve conformal
symmetry. The automorphism~$\Omega$ induces a permutation $\omega:
\RA\to\RA$ on the set of representations which leaves invariant the
vacuum representation (see e.g.\ 
\cite{Recknagel:1998sb} for details). It is then easy to see that 
for each element
\eq
  (\mu,\eta)\in\Spec^\omega \ =\ 
  \bigl\{\:(\mu,\eta)\:\bigl|\:(\mu,\bar{\mu}|\eta)
  \in\Spec\text{ and }\bar{\mu}=\omega(\mu^+)\bigr\}
\qe
in the $\omega$-symmetric part of the spectrum one can construct 
a so-called Ishibashi (or {\em generalized coherent}) state 
$|\mu,\eta\rrangle$. These states are normalized by 
\eq
  \llangle\mu,\eta|q^{\frac{1}{2}(L_0+\bar{L}_0-c/12)}|\nu,
  \epsilon\rrangle
  \ =\ \delta_\nu^\mu\,\delta_\epsilon^\eta\:\chi_\mu(q)
\qe
and they constitute a complete linear independent set of solutions
to the linear equations~\eqref{eq:BoundaryCondition}. Although the 
Ishibashi states are often said to live in the bulk Hilbert space 
$\Hilb$, one should bear in mind that they are not normalizable 
in the standard sense.
\smallskip%
  
Naively one could think that all linear combinations
\eqn
  \label{eq:BSDef}
  |b\rangle
  \ =\ \sum_{(\mu,\eta)\in\Spec^\omega}
  \frac{{\psi_b}^{(\mu,\eta)}}{\sqrt{S_{0\mu}}}\:|\mu,\eta\rrangle
\qen
would lead to consistent boundary states. There exists, however, 
the important {\em Cardy constraint} which arises from world-sheet 
duality or from an exchange of open and closed string channel in 
a more string theoretic language. More precisely, one has 
\eq
Z_{ab}(q)  & = &  
  \langle a|\tilde{q}^{\frac{1}{2}(L_0+\bar{L}_0-c/12)}|b\rangle
  \  = \ \sum_{(\mu,\eta)\in\Spec^\omega}
  \frac{{\bar{\psi_a}}^{(\mu,\eta)}{\psi_b}^{(\mu,\eta)}}{S_{0\mu}}
 \, \chi_\mu(\tilde{q})\nonumber \\[2mm] 
 & = & \hspace*{-5mm} \sum_{\substack{(\mu,\eta)\in\Spec^\omega\\[1mm] 
 \nu\in\RA}}
 \frac{{\bar{\psi_a}}^{(\mu,\eta)}{\psi_b}^{(\mu,\eta)}S_{\nu\mu}}
 {S_{0\mu}}\, \chi_\nu(q) \ \equiv \ \sum_{\nu\in\RA}\:{\bigl(n_\nu\bigr)_b}^a
 \:\chi_\nu(q)    
\qe
where~$\tilde{q}$ is obtained from~$q$ by modular transformation, 
i.e.\ $\tilde{q}=e^{-2\pi i/\tau}$. All characters $\chi_\nu(q)$ in
the second line must have non-negative integer coefficients
${\bigl(n_\nu\bigr)_b}^a$ since we want to interprete the whole expression 
as an open string partition function. Consistent boundary states which 
correspond to the gluing automorphism~$\Omega$ may be generated from a 
set~$\B^\omega$ of elementary boundary states. As a criterion for 
elementarity we shall use the requirement ${\bigl(n_0\bigr)_b}^a=
\delta_b^a$. It states that the identity field should only live 
between identical boundary conditions and that it should appear 
with multiplicity one. Every consistent boundary state may be 
represented as a superposition of elementary boundary states 
with non-negative integer coefficients. Let us emphasize that the boundary 
states do not only depend on the gluing automorphism~$\Omega$ but 
also on the bulk partition function under consideration.%
\smallskip%

One can show~\cite{Behrend:1999bn} that the matrices~$n_\nu$ form 
a non-negative integer valued matrix representation (NIM-rep) of 
the fusion ring of the CFT, i.e.
\eqn
  \label{eq:NIMrep}
  n_\lambda \, n_\mu \ =\ \sum_{\nu\in\RA}{N_{\lambda\mu}}^\nu \,
  n_\nu\qquad\text{ and } \qquad n_{\lambda^+}\ =\ (n_\lambda)^T
\qen
where the fusion rules of $\A$ are denoted by ${N_{\lambda\mu}}^\nu$.
Let us also remark that the classification of NIM-reps for a given 
fusion ring is not sufficient to construct consistent BCFTs. In fact, 
many NIM-reps are known to possess no physical interpretation
\cite{Gannon:2001ki}.
\medskip 

There is a class of boundary conditions which was constructed by 
Cardy more than ten years ago 
\cite{Cardy:1989ir}. In the original setup, these boundary conditions 
require that $\Omega$ is the identity and that we are working with 
the charge conjugated modular invariant, i.e.\  $Z^{\mu\bar\mu} = 
\delta^{\mu \mu^+}$. Hence, we can identify $\Spec^{\id}$ with 
$\RA$. It is then easy to solve the Cardy condition by the boundary 
states
\eqn
  \label{eq:CardyPartition}
  |\nu\rangle
  \ =\ \sum_{\lambda\in\RA}\frac{S_{\nu\lambda}}{\sqrt{S_{0\lambda}}}
  \:|\lambda\rrangle
\qen
where $\nu \in \B^{\id} \cong \RA$. Indeed, the Verlinde formula 
for fusion coefficients~\cite{Verlinde:1988sn}
\eqn
  \label{eq:VerlindeFormula}
  {N_{\mu\nu}}^\lambda
  \ =\ \sum_{\sigma\in\RA}\frac{\bar{S}_{\lambda\sigma}S_{\mu\sigma}
  S_{\nu\sigma}}{S_{0\sigma}}
\qen
  immediately implies
\eq
  Z_{\mu\nu}(q)
  \ =\ \sum_{\lambda\in\RA}\:{N_{\mu^+\nu}}^\lambda\:\chi_\lambda(q)
  \ \ .
\qe
For later convenience let us summarize some important properties of 
fusion rules which may easily be proved by means of the Verlinde 
formula~\eqref{eq:VerlindeFormula} using the properties 
\eqref{eq:SMatrixProperties} for the modular S-matrix,  
\eqn
  \label{eq:FusionProperties1}
  {N_{0\mu}}^\sigma\ =\ \delta_\mu^\sigma\ \ , \hspace{1cm}
   {N_{\mu\nu}}^{\sigma} & = & {N_{\nu\mu}}^{\sigma}
  \ =\ {N_{\mu\sigma^+}}^{\nu^+}\ =\ {N_{\mu^+\nu^+}}^{\sigma^+}
 \\[3mm] \label{eq:FusionProperties2}
  \sum_{\sigma \in \Rep(\A)}{N_{\lambda\nu}}^{\sigma}
   {N_{\mu\sigma}}^{\rho}
  & = & \sum_{\sigma \in \Rep(\A)}{N_{\mu\lambda}}^{\sigma}
    {N_{\sigma\nu}}^{\rho}  
\ \ .
\qen
The first equation guarantees that the identity field propagates 
only between identical boundary conditions.
In addition, these relations 
imply that the matrices ${(N_\lambda)_\mu}^\sigma = 
{N_{\lambda\mu}}^\sigma$ form a representation~\eqref{eq:NIMrep}
-- the adjoint representation -- of the fusion algebra.

\subsection{The coset construction}

One of the basic tools to build new conformal field theories is
the so-called coset or GKO construction~\cite{Goddard:1985vk}. 
Although the main ideas in this section and in the rest of the 
paper apply to a rather general class of coset theories, we shall 
specialize most of our presentation to affine Kac-Moody algebras
and their cosets. The formulation we have chosen, however, 
suggests the appropriate generalization. 
\smallskip   

Let $\G$ be a semi-simple simply connected compact group and 
$\ag_k$ the associated affine Kac-Moody algebra. The latter 
generates a chiral algebra that we denote by $\A(\G) = 
\A(\ag_k)$. Now we want to choose a semi-simple subgroup $\P$ 
of $\G$. Up to isomorphism, the embedding $\mf{p}\embin\mf{g}$ 
of the corresponding Lie algebras can be defined  by giving a 
projection
$\mc{P}:L_w^{(\g)}\to L_w^{(\p)}$ from the weight lattice of~$\g$ 
to the weight lattice of~$\p$. This projection is just dual to the 
injection of Cartan subalgebras. The embedding of Lie algebras may 
be lifted to an embedding $\ap_{k^\prime}\embin\ag_k$ of untwisted 
affine Kac-Moody algebras where the levels are related by 
$k^\prime=x_ek$ with embedding index $x_e$ (see~\cite{FrancescoCFT} 
for instance). By the GKO construction~\cite{Goddard:1985vk} one may
then define the coset chiral algebra $\A(\G/\P)$ such that the energy 
momentum tensors satisfy $T^{\G}=T^{\G/\P}+T^{\P}$ and all the chiral 
fields generating $\A(\G/\P) \subset \A(\G)$ commute with those in 
$\A(\P)$. It was shown in~\cite{Gawedzki:1988hq, Gawedzki:1989nj, 
Nahm:1987,Bardakci:1988ee} that the coset chiral algebra describes 
the symmetry of the $\G/\P$ gauged WZW model. There are (at least) 
two equivalent ways of analysing the coset chiral algebra. First, 
there is a more geometric one which is discussed for example in
\cite{FrancescoCFT}. In our presentation we will follow the simple 
current approach~\cite{Schellekens:1989am,Schellekens:1989dq,
Schellekens:1990xy,Intriligator:1990zw} as this 
allows a straightforward generalization to cosets which do not arise
from WZW theories.
\smallskip

We start with a discussion of simple currents $J \in \Rep(\G)=\Rep 
\bigl(\A(\G)\bigr)$  which are characterized by the property that the
fusion $(J)\star(\mu)$ of $J$ with any other sector $\mu\in\Rep(\G)$ 
contains exactly one representation. We shall denote the latter 
by $J\mu\in\Rep(\G)$. Since the vacuum representation is a simple 
current and the fusion product is commutative, the set of all 
simple currents forms an abelian group $\Cent(\G)$. In almost all
cases,\footnote{The only exception is $E_8$ at level 2.} this 
group is isomorphic to the center of the Lie group $\G$. This also 
means that $\Cent(\G)$ is in one-to-one correspondence with 
symmetries of the Dynkin diagram 
of the affine Lie algebra $\ag$ modulo those of the Dynkin diagram 
of $\g$.
\smallskip

Let us now summarize some well-known properties of simple currents.
It turns out that simple current transformations satisfy
\eqn
  \label{eq:SMatrixSymmetry}
  S_{J\mu\ \nu}^{\G}\ =\ e^{2\pi iQ_J(\nu)}\, S_{\mu\nu}^{\G}\ \ .
\qen
The number $Q_J(\nu)$ is defined modulo integers and it is called 
the {\em monodromy charge} of~$\nu$ with respect to~$J$. It is 
possible to show that monodromy charges are related to conformal 
weights by the formula
\eqn
  \label{eq:MonodromyCharge}
  Q_J(\nu)\ =\ h_J+h_\nu-h_{J\nu}\mod1\ \ .
\qen
The relation~\eqref{eq:SMatrixSymmetry} has some wide reaching 
consequences. In particular, iterated application implies
\eq
  Q_{J^n}(\nu)\ =\ nQ_{J}(\nu) \ \ \ , \ \ \ 
  Q_{J}(J^\prime\nu)+Q_{J^\prime}(\mu)\ =\ 
  Q_{J^\prime}(J\mu)+Q_{J}(\nu)\ \ .
\qe
For a simple current $J$ of order $N$, i.e.\ an element $J \in 
\Cent (G)$ satisfying $J^N=(0)$ (the vacuum representation), the 
first relation means that the monodromy charge $\exp\bigl(2\pi i 
Q_J(\nu)\bigr)$ is an $N^{th}$ root of unity. Simple currents  
provide symmetries of the fusion rules. Indeed, the S-matrix 
symmetry~\eqref{eq:SMatrixSymmetry} in combination with the 
Verlinde formula~\eqref{eq:VerlindeFormula} implies
\eqn
  \label{eq:FusionSymmetry}
  {N_{\mu J\nu}}^{J\sigma}\ =\ {N_{\mu\nu}}^\sigma\ \ .
\qen
With this preparation on simple currents we can now address 
our main aim to describe properties of coset chiral algebras. 
\medskip 

It is convenient to distinguish the sectors of the $\G$ and $\P$ 
theories by using different types of labels, 
\eq
  \ag_k: \:\mu,\nu,\rho,\ldots\in\Rep(\G)\qquad\qquad
  \ap_{k^\prime}: \: a,b,c,\ldots\in\Rep(\P)\ \ .
\qe
The generic coset sectors are labeled by tupels $(\mu,a)$
satisfying some constraints which are known as branching 
selection rules and depend on the specific algebras and 
embeddings. To be concrete, any allowed pair $(\mu,a)$ has 
to satisfy
\eqn
  \label{eq:BranchingSelectionG}
  \mc{P}\mu-a\ \in\ \mc{P}\mc{Q}
\qen
where~$\mc{Q}$ denotes the root lattice of~$\g$. If this relation 
would not be satisfied, there would be no chance to find a 
weight in the weight system of~$\mu$ that is projected 
onto~$a$. We denote the set of allowed coset labels by 
\eq
  \Allowed(\G/\P) \ =\ \bigl\{\,(\mu,a)\,\bigl|\,\mc{P}\mu-
  a\in\mc{P}\mc{Q}\,\bigr\} \ \subset\ \Rep(\G)\times\Rep(\P)\ \ . 
\qe
In addition, certain pairs $(\mu,a)$ need to be identified
because they give rise to one and the same sector 
\cite{Gepner:1989jq,Moore:1989yh}. Generically, this field
identification 
exactly corresponds to elements in the common center
$\Cent(\G)\cap\Cent(\P)$ of the groups $\G$ and $\P$.\footnote{In 
so-called Maverick cosets (see e.g.\ \cite{Dunbar:1993hr}) and 
cosets arising from conformal embeddings this statement is not 
true. Conformal embeddings, however, are restricted to $k=1$ and 
all known Maverick cosets also are a low level phenomenon.}
\smallskip

Before we continue, let us make the last statement precise.
To this end, we pick two elements $J\in\Cent(\G)$ and $J^\prime
\in\Cent(\P)$. We say that the pair $(J,J^\prime)$ lies in the 
common center if the relation
\eqn
  \label{eq:BranchingAutomorphisms}
  Q_J(\mu)\ =\ Q_{J^\prime}(\mc{P}\mu)
\qen
holds for all weights $\mu \in \Rep(\G)$. The abelian group of all 
pairs $(J,J^\prime)$ satisfying this condition shall be denoted by
$\Gid$. By construction, $\Gid$ is a subgroup of the product
$\Cent(\G)\times\Cent(\P)$. Sometimes this group also is 
called identification group. Note that it depends explicitly 
on the embedding. 
\smallskip

The first application of this identification group $\Gid$ is
that it allows to reformulate the branching selection rule 
\eqref{eq:BranchingSelectionG} in a completely algebraic way. 
It turns out that the allowed weights may be described by
\eqn
  \label{eq:BranchingSelection}
  \Allowed(\G/\P)
  \ =\ \bigl\{\,(\mu,a)\,\bigl|\, Q_J(\mu)=Q_{J^\prime}(a)\text{ for all }
  (J,J^\prime)\in\Gid\,\bigr\}\ \ . 
\qen
Moreover, we can now also address the issue of field identification. 
Let us note that for generic sectors $(\mu,a),(\nu,b) \in 
\Allowed(\G/\P)$, the modular S-matrix of the coset theory is 
given by
\eqn
  \label{eq:CosetSMatrix}
  S_{(\mu,a)(\nu,b)}^{\G/\P}
  &=&|\Gid|\:S_{\mu\nu}^{\G}\ \bar{S}_{ab}^{\P}\ \ .
\qen
If we act on the first weight by an element $(J,J^\prime)\in\Gid$ we
obtain
\eq
  S_{(J\mu,J^\prime a)(\nu,b)}^{\G/\P}
  &=&e^{2\pi i (Q_J(\nu)-Q_{J^\prime}(b))}
 \:S_{(\mu,a)(\nu,b)}^{\G/\P}
  \:=\:S_{(\mu,a)(\nu,b)}^{\G/\P}
\qe
from equation~\eqref{eq:SMatrixSymmetry}. The phase factor vanishes 
because of the branching selection rule expressed in the 
relation~\eqref{eq:BranchingSelection}. This hints towards an 
identification of the sectors $(J\mu,J^\prime a)$ and $(\mu,a)$.  
In fact, under certain simplifying assumptions one can show that 
inequivalent irreducible representations of the coset theory are
labeled by elements  
\eqn
  \label{eq:CosetRepresentations}
  [\mu,a]\in\Rep(\G/\P)\ =\ \Allowed(\G/\P)/\Gid\ \ .
\qen
Complications arise when there exist fixed points, i.e.\ sectors 
with the property $(J\mu,J^\prime a)=(\mu,a)$ for at least one 
pair $(J,J^\prime)\in\Gid$. In this case, the representation spaces
carry (reducible) representations of the relevant stabilizer 
subgroup of $\Gid$. Determining the irreducible constituents 
and the associated modular data is known as fixed point 
resolution~\cite{Schellekens:1990uf,Schellekens:1990xy,
Fuchs:1996zr,Fuchs:1996tq}. We will circumvent these technical 
difficulties and assume in the following that our field 
identification has no fixed points. Under these circumstances
all orbits $\Gid(\mu,a)$ have the same length $|\Gid|$.
\smallskip

From the expression (\ref{eq:CosetSMatrix}) and the Verlinde 
formula we can easily deduce the following expression for 
the fusion coefficients of the coset model,  
\eqn \label{cosN} 
N_{[\mu,a],[\nu,b]}^{[\sigma,c]}\ =\ \sum_{(J,J^\prime) \in\Gid}\ 
N_{\mu\nu}^{J\sigma}\ N_{ab}^{J^\prime c}\ \ .
\qen
Below we shall also need a projector which implements the branching 
selection rule~\eqref{eq:BranchingSelection}. This is rather 
easy to introduce by the explicit formula 
\eqn
  \label{eq:Projector}
  P(\mu,a)
  \ =\ \frac{1}{|\Gid|}\sum_{(J,J^\prime)\in\Gid}
  e^{2\pi i (Q_{J}(\mu)-Q_{J^\prime}(a))}\ \ .
\qen
The definition of $\Allowed(\G/\P)$ directly implies that $P(\mu,a) = 1$ for 
all $(\mu,a)$ in the set $\Allowed(\G/\P)$ and that it vanishes 
otherwise. 

\section{\label{sc:NewStates}The new boundary states} 
 
Our aim now is to construct new boundary states for a theory whose  
partition function is given by the charge conjugated modular invariant 
of the chiral algebra $\A(\G)$. We shall analyse this theory with 
respect to some intermediate chiral algebra $\A(\P)$. This will lead 
us to a set of boundary conditions extending the usual Cardy type
conditions. Explicit expressions for the boundary states and the 
associated open string spectra are provided for different gluing 
conditions.     

\subsection{Decomposition of the bulk modular invariant} 
 
Before going into the discussion of the boundary states, let us 
present the general idea of our construction. As usual, our starting 
point is some bulk theory with a state space that is assumed to be 
charge conjugated with respect to some chiral algebra $\A(\G)$, 
\eq 
  \Hilb^{\G} 
  \ \ \cong\ \bigoplus_{\mu\in\Rep(\G)} 
   \Hilb_\mu^{\G}\otimes\Hilbb_{\mu^+}^{\G}\ \ . 
\qe 
But now we want to construct boundary states which break at least 
some part of the chiral symmetry. To be precise, we only want to 
preserve the subalgebra 
\eq 
  \A(\G/\P)\oplus \A(\P)\ \embin\ \A(\G) 
\qe 
at the boundary. It is then natural to decompose the full state 
space according to the action of the smaller chiral algebra. Under 
the restriction to $\A(\G/\P)\oplus \A(\P)$, the irreducible 
representations of $\A(\G)$ can be reduced to 
\eq 
  \Hilb_{\mu}^{\G} \ \
  \cong  \bigoplus_{(\mu,a)\in\Allowed(\G/\P)}\: 
  \Hilb_{(\mu,a)}^{\G/\P}\otimes\Hilb_a^{\P}\ \ . 
\qe 
Note that the sum is restricted to those values of $a$ for which the branching 
selection rule~\eqref{eq:BranchingSelection} is satisfied. The last relation 
also illustrates why we wanted to preserve the chiral algebra $\A(\G/\P)
\oplus \A(\P)$, not only the subalgebra $\A(\P)$: as representations of
$\A(\G/\P)$ are generically infinite-dimensional the resulting theory 
would become non-rational otherwise.
The decomposition of the full space reads 
\eqn 
  \label{eq:HilbertDecomposed} 
  \Hilb^{\G} 
  \ \cong\ \bigoplus_{(\mu,a),(\mu,\bar{a})\in\Allowed(\G/\P)}\: 
  \Hilb_{(\mu,a)}^{\G/\P}\otimes\Hilb_a^{\P}\otimes
  \Hilbb_{(\mu,\bar{a})^+}^{\G/\P}\otimes\Hilbb_{\bar{a}^+}^{\P}\ \ . 
\qen 
In terms of partition functions, the decomposition can be expressed 
as follows, 
\eqn \label{GPF} 
Z \ = \ \sum_{\mu\in\RA} \Bigl| \, \chi^{\G}_\mu \, \Bigr|^2 \ = \ 
 \sum_{\mu\in\RA}\,\biggl| \, \sum_{(\mu,a)\in\Allowed(\G/\P)} \chi^{\G/\P}_{(\mu,a)}\:\chi^\P_a 
  \, \biggr|^2 \ \ .
\qen 
To simplify notations we have used that the 
characters are invariant under the substitution $\mu \mapsto 
\mu^+$. Hence, on the level of partition functions, we do not 
distinguish between the diagonal and the charge conjugated 
modular invariant. Let us stress that the theory is not charge 
conjugated with respect to the smaller chiral algebra. In particular, 
the boundary states preserving the smaller chiral algebra can not be 
constructed by Cardy's solution. 
\smallskip

In our setting we are free to choose two different gluing automorphisms 
$\Omega^{\G/\P}$ and $\Omega^{\P}$ for chiral fields in the two
individual parts of the reduced chiral algebra and to require   
\eq 
  \bigl(\phi(z)-\Omega^{\G/\P}\bar{\phi}(\bar{z})\bigr)|B\rangle&=&0
  \\[2mm] 
  \bigl(\psi(z)-\Omega^{\P}\bar{\psi}(\bar{z})\bigr)|B\rangle&=&0 
\qe 
for arbitrary fields $\phi\in\A(\G/\P)$ and $\psi\in\A(\P)$. Note that these
conditions ensure the Virasoro field $T^\G = T^{\G/\P} + T^\P$ of the 
theory to be preserved along the boundary. Naively one might think that 
boundary states satisfying these gluing conditions can be factorized 
into boundary states of the two chiral algebras $\A(\G/\P)$ and $\A(\P)$. 
However, this is not true because the partition function does not 
factorize. 
\smallskip

We will certainly not be able to solve the boundary theories for an 
arbitrary choice of $\Omega^{\G/\P}$ and $\Omega^{\P}$. In the next 
subsection we shall discuss the special case in which both these 
gluing automorphisms are trivial. After that, we address a more 
general possibility in which $\Omega^{\G/\P}$ is still trivial 
while any choice of $\Omega^{\P}$ is allowed. 
 
\subsection{\label{sc:TrivialTrivial}Trivial gluing automorphisms} 
 
We start with boundary conditions for which the left and right movers 
are glued trivially, $\Omega= \Omega^{\G/\P}\otimes \Omega^{\P} = 
\id\otimes\id$. This induces the identity map $\omega=\id\times\id$ 
on the set $\Rep(\G/\P)\times\Rep(\P)$ of sectors. The constituents of 
the Hilbert space $\Hilb^{\G}$ which are left-right-symmetric with 
respect to the automorphism~$\omega$ are given by 
\eq 
  \Hilb_{(\mu,a)}^{\G/\P}\otimes\Hilb_a^{\P}\otimes
  \Hilbb_{(\mu,a)^+}^{\G/\P}\otimes\Hilbb_{a^+}^{\P}\ \ . 
\qe 
Hence, Ishibashi states are labeled unambigously by pairs 
$(\mu,a)\in\Allowed(\G/\P)$, i.e. $\mu,a$ run over all representations such 
that the branching selection rule~\eqref{eq:BranchingSelection} is satisfied. 
Let us point out that in these labels, no field identification is made. We 
choose the standard normalization~\cite{Behrend:1999bn} of Ishibashi states 
such that 
 \eq 
  \llangle(\mu,a)|\,\tilde{q}^{\frac{1}{2}(L_0+\bar{L}_0-c/12)}\,
     |(\nu,b)\rrangle 
   &=&\delta_\nu^\mu\, \delta_b^a\ \chi_{(\mu,a)}^{\G/\P}(\tilde{q})
  \, \chi_a^{\P}(\tilde{q})\ \ . 
\qe 
As we shall see, the elementary boundary states are labeled by 
elements $(\rho,r)$ from the set  $\B^{\id\times\id}=\bigl(\Rep(\G)
\times \Rep(\P)\bigr)/\Gid$. Their expansion in terms of Ishibashi 
states reads 
\eqn \label{sol1a} 
  |(\rho,r)\rangle 
  \ =\ \sum_{(\mu,a)\in\Allowed(\G/\P)}\, B_{(\rho,r)}^{(\mu,a)}\,
       |(\mu,a)\rrangle 
\qen
with coefficients $B_{(\rho,r)}^{(\mu,a)}$ being determined by the
modular S-matrix of the $\G$ and the $\P$ theory through the simple formula 
\eqn \label{sol1b}
  B_{(\rho,r)}^{(\mu,a)} 
  \ =\ \frac{S_{\rho\mu}^{\G}}{\sqrt{S_{0\mu}^{\G}}}\:
  \frac{\bar{S}_{ra}^{\P}}{\bar{S}_{0a}^{\P}} \ \ . 
\qen    
The proof of this claim proceeds in several steps.
Let us first note that $(\rho,r)$ and $(J\rho,J^\prime r)$ lead to the
same boundary state. This is a simple consequence
of eq.~\eqref{eq:SMatrixSymmetry} and the
definition~\eqref{eq:BranchingSelection}
of $\Allowed(\G/\P)$. We will show now that 
the proposed boundary states possess a consistent open string spectrum. 
Finally, it remains
to demonstrate that the identity field propagates in between two 
boundary conditions if and only if these two boundary conditions 
are identical. 
\medskip 

Let us begin by computing the open string spectrum in between two 
boundary conditions $(\rho_1,r_1)$ and $(\rho_2,r_2)$,\footnote{To save 
space we omit the ranges of the summation indices. The summation 
rules are: $(\mu,a)\in\Allowed(\G/\P)$, $[\mu,a]\in\Rep(\G/\P)$ and 
all other (single) indices run over $\Rep(\G)$ or $\Rep(\P)$,  
respectively.}  
\eq 
  Z & = & Z_{(\rho_1,r_1),(\rho_2,r_2)}(q) 
  \ =\  \  \langle(\rho_1,r_1)|\,\tilde{q}^{\frac{1}{2}
  (L_0+\bar{L}_0-c/12)}\,|(\rho_2,r_2)\rangle\\[2mm]  
  &=&\sum_{(\mu,a),[\nu,b],c}\:\Bigl[\:\bar{B}_{(\rho_1,r_1)}^{(\mu,a)}
  B_{(\rho_2,r_2)}^{(\mu,a)}S_{(\mu,a),(\nu,b)}^{\G/\P}S_{ac}^{\P}
  \:\Bigr]\:\chi_{(\nu,b)}^{\G/\P}(q)\, \chi_c^{\P}(q) \\[2mm]
  &=&|\Gid| 
  \sum_{(\mu,a),[\nu,b],c}\:\Biggl[\:\frac{\bar{S}_{\rho_1\mu}^{\G}
  S_{\rho_2\mu}^{\G}S_{\nu\mu}^{\G}}{S_{0\mu}^{\G}}\:
  \frac{S_{r_1a}^{\P}\bar{S}_{r_2a}^{\P}\bar{S}_{ba}^{\P}
  S_{ca}^{\P}}{S_{0a}^{\P}\bar{S}_{0a}^{\P}}\:\Biggr]\:
  \chi_{(\nu,b)}^{\G/\P}(q)\,\chi_c^{\P}(q)\ \ . 
\qe 
In the second step we inserted our expression for the coefficients
of the boundary states and formula (\ref{eq:CosetSMatrix}) for the 
S-matrix of the coset model. Note that the coefficients of the 
individiual characters on the right hand side are not expected 
to be integers since we still sum over labels which are related by 
the action of the identification group. Now we use that the quantum 
dimensions $S_{ra}/S_{0a}$ form a representation of the fusion algebra,  
\eq 
  \frac{S_{r_1a}^{\P}}{S_{0a}^{\P}}\:\frac{\bar{S}_{r_2a}^{\P}}
  {\bar{S}_{0a}^{\P}} &=&\sum_{d\in\Rep(\P)}{N_{r_1r_2^+}}^d\
  \frac{S_{da}^{\P}}{S_{0a}^{\P}}\ \ ,  
\qe 
and obtain 
\eq 
  Z&=&|\Gid| 
  \sum_{(\mu,a),[\nu,b],c,d}\:{N_{r_1r_2^+}}^d\: 
  \Biggl[\:\frac{\bar{S}_{\rho_1\mu}^{\G}S_{\rho_2\mu}^{\G}
   S_{\nu\mu}^{\G}}{S_{0\mu}^{\G}}\:\frac{S_{da}^{\P}
   \bar{S}_{ba}^{\P}S_{ca}^{\P}}{S_{0a}^{\P}}\:\Biggr]\:
   \chi_{(\nu,b)}^{\G/\P}(q)\,\chi_c^{\P}(q)\ \ . 
\qe 
If the sum over the pairs $(\mu,a)$ was not restricted 
by the branching selection rule~\eqref{eq:BranchingSelection}, 
the quotients of S-matrices could be evaluated by means of the 
Verlinde formula~\eqref{eq:VerlindeFormula}. But as it stands, 
this step can not be performed so easily. However, we can 
implement the constraint my means of the projector $P(\mu,a)$ 
which has been defined in~\eqref{eq:Projector}. This yields 
\eq 
  Z&=&|\Gid| 
  \sum_{\mu,a,[\nu,b],c,d}\:P(\mu,a)\:{N_{r_1r_2^+}}^d\: 
  \Biggl[\:\frac{\bar{S}_{\rho_1\mu}^{\G}S_{\rho_2\mu}^{\G}
   S_{\nu\mu}^{\G}}{S_{0\mu}^{\G}}\:\frac{S_{da}^{\P}
   \bar{S}_{ba}^{\P}S_{ca}^{\P}}{S_{0a}^{\P}}\:\Biggr]\:
   \chi_{(\nu,b)}^{\G/\P}(q)\,\chi_c^{\P}(q)\\ 
  &=&\sum_{\substack{\mu,a,[\nu,b],c,d\\(J,J^\prime)\in\Gid}}
  \frac{e^{2\pi i Q_{J}(\mu)}}{e^{2\pi i Q_{J'}(a)}}
  \:{N_{r_1r_2^+}}^d\: 
  \Biggl[\:\frac{\bar{S}_{\rho_1\mu}^{\G}S_{\rho_2\mu}^{\G}
   S_{\nu\mu}^{\G}}{S_{0\mu}^{\G}}\:\frac{S_{da}^{\P}
   \bar{S}_{ba}^{\P}S_{ca}^{\P}}{S_{0a}^{\P}}\:\Biggr]\:
   \chi_{(\nu,b)}^{\G/\P}(q)\,\chi_c^{\P}(q)\ \ .  
\qe 
We then use the fact that the exponentials may be pulled into 
the S-matrices with the help of eq.~\eqref{eq:SMatrixSymmetry}. 
The result is 
\eq 
  Z&=&\sum_{\substack{\mu,a,[\nu,b],c,d\\(J,J^\prime)\in\Gid}}
  \:{N_{r_1r_2^+}}^d\: 
  \Biggl[\:\frac{\bar{S}_{\rho_1\mu}^{\G}S_{\rho_2\mu}^{\G}
  S_{J\nu\ \mu}^{\G}}{S_{0\mu}^{\G}}\:\frac{S_{da}^{\P}
  \bar{S}_{J^\prime b\ a}^{\P}S_{ca}^{\P}}{S_{0a}^{\P}}\:\Biggr]\:
  \chi_{(\nu,b)}^{\G/\P}(q)\,\chi_c^{\P}(q)\ \ . 
\qe 
As the field identification demands $\chi_{(\nu,b)}^{\G/\P}=
\chi_{(J\nu,J^\prime b)}^{\G/\P}$, we may collect the summations 
over $(J,J^\prime )\in\Gid$ and $[\nu,b]\in\Rep(\G/\P)$ to give a 
sum over $(\nu,b)\in\Allowed(\G/\P)$. Then, applying in addition 
the Verlinde formula~\eqref{eq:VerlindeFormula}, we finally 
arrive at 
\eqn 
  \label{eq:CosetOpenPartition} 
  Z_{(\rho_1,r_1),(\rho_2,r_2)}&=& 
  \sum_{\substack{(\nu,b)\in\Allowed(\G/\P)\\c,d\in\Rep(\P)}}\:
  {N_{\rho_1^+\rho_2}}^{\nu}\,{N_{r_1^+r_2}}^d\,{N_{dc}}^{b}\:
  \:\chi_{(\nu,b)}^{\G/\P}(q)\,\chi_c^{\P}(q)\ \ . 
\qen 
In the last step we also used some symmetries of the fusion rule 
coefficients~\eqref{eq:FusionProperties1} and charge conjugation invariance 
of the characters. Thereby we have shown that $Z$ can be expanded into 
characters of the chiral algebra $\A(\G/\P) \oplus \A(\P)$ with 
manifestly non-negative integer coefficients. 
\smallskip

Finally, we now wish to convince ourselves that the vacuum 
representation appears exactly once in the boundary partition 
function~\eqref{eq:CosetOpenPartition} with two identical elementary 
boundary conditions and that it does not contribute whenever the two 
boundary conditions are different. This is somewhat obscured by the 
possible field identification. By a calculation similar to the 
previous one it is possible to rewrite the partition function 
in the form 
\eq 
  Z_{(\rho_1,r_1),(\rho_2,r_2)} 
  &=&\sum_{(J,J^\prime )\in\Gid}\:\delta_{J\rho_2}^{\rho_1} 
     \delta_{J^\prime r_2}^{r_1}\:\chi_{(J0,J^\prime 0)}^{\G/\P}(q)
     \,\chi_0^{\P}(q) 
     \:+ \: \dots 
\qe 
where $\dots$ stand for other contributions that do not contain the 
vacuum character of $\A(\G/\P) \oplus \A(\P)$. Hence, the  
identity field only appears if $(\rho_1,r_1)$ and $(\rho_2,r_2)$ 
are identical up to the action of $\Gid$, in agreement with our 
claim. 
\medskip 
 
Let us conclude with some comments. First note that the partition 
function~\eqref{eq:CosetOpenPartition} can not be obtained by decomposing 
the usual Cardy boundary partition function. In fact, if we decompose the 
latter into characters of $\A(\G/\P) \oplus \A(\P)$, we obtain 
\eq 
  Z_{\rho_1\rho_2} 
  \ =\ \sum_{\nu\in\Rep(\G)}\:{N_{\rho_1^+\rho_2}}^\nu\:\chi_{\nu}^{\G}(q) 
  \ = \ 
  \:\sum_{(\nu,b)\in\Allowed(\G/\P)}\: 
     {N_{\rho_1^+\rho_2}}^\nu\:\chi_{(\nu,b)}^{\G/\P}(q)\,\chi_b^{\P}(q)  
\qe 
where the labels $b$ for the $\A(\P)$-sector coincide with the second 
label of the coset theory. But this is not the case for most of the 
partition functions of our boundary states. In other words, our new 
boundary theories manifestly break some of the chiral symmetry $\A(\G)$ 
in the bulk theory. Note, however, that the right hand side of the 
previous equation coincides with the partition function for the pair 
$(\rho_1,0), (\rho_2,0)$. In other words, for the states of the special 
form $|(\rho,0)\rangle$, the maximal chiral symmetry is restored and 
these states can be identified with Cardy's boundary states. 
 
\subsection{\label{sc:TrivialTwisted}The case of partially twisted 
boundary conditions} 
 
Our construction possesses a natural extension to cases in which 
we choose a non-trivial gluing automorphism for one of the factors 
$\A(\G/\P)$ or $\A(\P)$. To be specific, we shall assume that the 
gluing automorphism $\Omega^{\G/\P}$ remains trivial. A solution is 
then possible for any $\Omega^\P$, provided we can solve the 
corresponding $\P$-theory with charge conjugated modular invariant 
partition function. For technical reasons we shall also assume that 
the identification group $\Gid$ is trivial. 
\smallskip 

To begin with, let us briefly comment on the solution of the auxiliary 
$\P$-theory with charge conjugate modular invariant partition function. 
As usual, boundary states for the gluing automorphism $\Omega^\P$ are 
built up from Ishibashi states $|a\rrangle^\P$ where $a \in \Rep(\P)$ 
and  $a  = \omega(a) := \omega^\P (a)$, 
\eq 
  |\alpha\rangle^{\P} 
   \ =\ \sum_{\omega(a)=a}
  \frac{{\psi_\alpha}^a}{\sqrt{S_{0a}^{\P}}}\:|a\rrangle^{\P}\ \ . 
\qe 
Here, $\a$ is chosen from the set $\a \in \B^\omega(\P)$. We will assume 
that the structure constants~${\psi_\alpha}^a$ are known explicitly. 
According to the general theory, they determine a NIM-rep through 
\eq 
  {\bigl(n_b\bigr)_\beta}^\alpha 
  \ =\ \sum_{\omega(a) = a}\frac{{\bar{\psi}_\alpha}^{\ a}\ {\psi_\beta}^a 
\ S_{ba}^\P}{S_{0a}^\P}\ \ . 
\qe 
For detailed expressions we refer the reader to 
\cite{Fuchs:1999zi,Birke:1999ik,Behrend:1999bn,Petkova:2002yj,
Gaberdiel:2002qa,Quella:2002wi} 
(see also~\cite{Alekseev:2002rj,Quella:2001wh} 
for the limit $k\to\infty$). 
\smallskip  
  
With this solution of the auxiliary problem in mind, we can 
return to our main goal of finding new symmetry breaking boundary 
states for the $\G$-theory. Once more, we have to determine which 
sectors in the decomposition~\eqref{eq:HilbertDecomposed} can 
contribute Ishibashi states. The condition is  
\eq 
  \bigl((\mu,\bar{a}),\bar{a}\bigr)^+ \ \sim\ 
  \bigl((\mu,a),\omega(a)\bigr)^+ \ \ . 
\qe 
We did not write ``$=$'' because the labels must be related only up  
to a field identification in the coset part. Since the $\Rep(\P)$ 
part is not subject to any field identification, the previous relation 
immediately implies~$\bar{a}=\omega(a)$. Hence we are left to decide 
whether for given $(\mu,a)\in\Allowed(\G/\P)$ we are able to find an 
element $(J,J^\prime )\in\Gid$ of the field identification group 
such that 
\eq 
  (J\mu,J^\prime a)\ =\ \bigl(\mu,\omega(a)\bigr)\ \ . 
\qe 
Up to now, we do not know how to determine all solutions to these 
equations in a systematic way. Obviously, such a classification 
depends strongly on the detailed structure of the field identification 
group and of its compatibility with the automorphism~$\omega$. In 
addition, it seems likely that one runs into troubles with field 
identification fixed points in the general case, a technical 
difficulty which we want to avoid.

We thus restrict ourselves to 
classes of embeddings for which the field identification group is 
trivial, i.e.\ $\Gid=\{\id\}$. This assumption in turn implies that 
there exist no branching selection rules. In particular, the coset 
representations are given by the set $\Rep(\G/\P)=\Rep(\G)\times
\Rep(\P)$. Let us emphasize that there is a large set of coset models 
for which our assumption holds, including all theories with an $E_8$ 
subgroup (which has trivial center) and the maximal embeddings 
$A_{n-1}\hookrightarrow A_n$ at embedding index $x_e=1$.
\smallskip

With our assumption $\Gid = \{\id\}$ being made, the decomposition
of the Hilbert space takes the particularly simple form 
\eqn 
  \label{eq:HilbertDecomposedPTwist} 
  \Hilb^{\G} 
  \ \cong \ \bigoplus_{\substack{\mu\in\Rep(\G)\\a,\bar{a}\in\Rep(\P)}}\: 
  \Hilb_{(\mu,a)}^{\G/\P}\otimes\Hilb_a^{\P}\otimes\ \Hilbb_{(\mu,\bar{a})^+}^{\G/\P}\otimes\Hilbb_{\bar{a}^+}^{\P}\ \ . 
\qen 
Ishibashi states in this case are given by $|(\mu,a),a\rrangle$ where $\mu
\in\Rep(\G)$ and $a\in\Rep(\P)$ with $\omega(a)=a$. Using the coefficients 
${\psi_\a}^a$ from the solution of the auxiliary $\P$-theory, we define 
boundary states by 
\eqn \label{sol2}  
  |(\rho,\gamma)\rrangle 
  \ =\ \sum_{\substack{\mu\in\Rep(\G)\\\omega(a)=a}}\:
  \frac{S_{\rho\mu}^\G}{\sqrt{S_{0\mu}^\G}}\:\frac{{{\bar{\psi}}_\gamma}^{\ a}}
   {\bar{S}_{0a}^\P} 
   \:|(\mu,a),a\rrangle\ \ . 
\qen 
Note that the expression imitates the construction of the last subsection. 
Along the line of our previous computations, one can also work out the 
boundary partition function. It is given by the formula
\eq 
  Z_{(\rho_1,\gamma_1),(\rho_2,\gamma_2)}^{\id\times\omega}(q) 
  \ =\ \sum_{\substack{\nu\in\Rep(\G)\\d\in\Rep(\P)}} 
   {N_{\rho_1^+\rho_2}}^{\nu}\:{N_{b^+c}}^{d}\:
   {\bigl(n_{d}\bigr)_{\gamma_1}}^{\gamma_2}\  
   \chi_{(\nu,b)}^{\G/\P}(q)\, \chi_c^{\P}(q) 
\qe 
which contains the NIM-rep that comes with our solution of the 
$\P$-theory. It is easy to check that this expression satisfies 
all consistency requirements.
  
Let us briefly comment on the generalization of this result to the case with
non-trivial field identification group $\Gid\neq\{\id\}$. The construction
of the previous section suggests that one needs properties of the coefficients
${\psi_\alpha}^a$ which are similar to those for the S-matrix given in
eq.~\eqref{eq:SMatrixSymmetry}. Such relations, however, have been worked out
in~\cite{Ishikawa:2001zu} for a number of examples.
 
\newpage
\section{Orbifold construction and brane geometry} 

In \cite{Maldacena:2001ky,Maldacena:2001xj} Maldacena, 
Moore and Seiberg developed a construction of symmetry 
breaking branes on group manifolds that is based on an orbifolding.
We shall now compare the results of their proposal with our 
algebraic analysis of symmetry breaking boundary states. After 
presenting the general ideas of the orbifold construction in the 
first subsection we show that it is capable of reproducing only 
a subset of our boundary states, namely those that are obtained 
by choosing an abelian denominator $\P$, i.e.\ $\Rep(\P) = 
\Cent(\P)$. Under this condition, our new boundary states possess 
a simple geometrical interpretation which emerges as a by-product 
of our discussion.

\subsection{\label{sc:Geometry} The orbifold construction - a review} 

Our aim here is to study branes in a simple current orbifold
of $\G/\P \times \P$. Before we address this rather complicated 
background, let us make some introductory remarks on brane 
geometries in group manifolds, cosets and orbifolds. As we 
have mentioned before, the description of branes in simple 
simply-connected compact group manifolds $\G$ involves the WZW 
theory  based on an affine Kac-Moody algebra $\ag_k$ with {\em 
charge conjugation} bulk partition function. The value of the 
level~$k$ controls the size of the group manifold. It is 
well-known that maximally symmetric D-branes on group manifolds 
are localized along quantized (twisted) conjugacy classes 
$$ 
  \mc{C}_\mu^\Omega \ =\  \bigl\{\, gh_\mu\Omega(g)^{-1}\, \bigl|\, g\ \in\ 
\G\, \bigr\}
$$
where $\Omega$ can be any automorphism of the group
$\G$~\cite{Alekseev:1998mc,Felder:1999ka}. Even though some of these
D-branes wrap trivial cycles, 
they are all stable due to the presence of a non-vanishing three form 
flux~\cite{Bachas:2000ik,Pawelczyk:2000ah,Bordalo:2001ec}. 
\smallskip 
  
A large variety of backgrounds arise from WZW models as cosets $\G/\P$ 
and orbifolds of the form $\G/\Gamma$. Maximally symmetric branes in 
coset theories with chiral algebra $\A(\G/\P)$ and charge conjugation 
bulk partition function are localized along the image of $\mc{C}_\mu^\Omega
\bigl(\mc{C}_a^\Omega\bigr)^{-1}$ under the projection from $\G$ to $\G/\P$ 
\cite{Maldacena:2001ky,Gawedzki:2001ye,Elitzur:2001qd,Fredenhagen:2001kw}. 
D-branes on orbifolds $\G/\Gamma$ can be represented by summing over all 
their pre-images in $\G$, at least as long as they do not contain fixed 
points for the action of $\Gamma$ on $\G$.  
\medskip

After this preparation, we now want to look at orbifolds obtained 
from product geometries of the form $\G/\P \times \P$. 
We request the orbifold group $\Gamma$ to be generated by simple 
currents, i.e.\ $\Gamma \subset \Cent(\G/\P) \times \Cent(\P)$. 
The sectors $\bigl([\mu,a],b\bigr)$ of $\A(\G/\P) \oplus \A(\P)$ 
fall into orbits $\bigl[[\mu,a],b\bigr]$ with respect to the 
action of $\Gamma$. With each of these orbits we associate two 
numbers, namely the monodromy charges $Q_{J}\bigl(\bigl[[\mu,a],
b\bigr]\bigr), J \in \Gamma,$ and the order $|\mc{S}_{[[\mu,a],b]}|$ 
of the stabilizer subgroup. The orbifold bulk partition function is 
then given by (see e.g.\ \cite{Schellekens:1990xy}) 
\eqn 
  \label{eq:OrbifoldPartition} 
  Z^{\text{orb}}(q) 
  \ = \sum_{Q_\Gamma([[\mu,a],b])=0}\Bigl|\mc{S}_{[[\mu,a],b]}\Bigr|\: 
   \Biggl|\sum_{([\nu,c],d)\in[[\mu,a],b]} 
   \chi_{(\nu,c)}^{\G/\P}(q)\chi_d^{\P}(q)\Biggr|^2\ \ . 
\qen
Boundary states of the orbifold theory can be obtained from the Cardy states 
$|[\mu,a],b\rangle=|[\mu,a]\rangle^{\G/\P}\otimes|b\rangle^{\P}$ of the charge 
conjugated covering theory by averaging over the action of the orbifold group 
$\Gamma$. This leads to boundary states of the form (see e.g.\ 
\cite{Brunner:2000nk})  
\eqn \label{orbbs} 
  \bigl|\bigl[[\mu,a],b\bigr]\bigr\rangle 
  \ = \ \frac{1}{\sqrt{|\Gamma|}}\sum_{(J,J^\prime)\in\Gamma} 
   |J[\mu,a],J^\prime b\rangle 
\qen
where the labels $\bigl[[\mu,a],b\bigr]$ of boundary states now take values
in the set $\bigl(\Rep(\G/\P)\times\Rep(\P)\bigr)/\Gamma$. The geometric 
interpretation of these boundary states is obvious from our remarks above.
It is also easy to calculate the boundary partition function 
\eq 
  Z_{[[\mu,a],b],[[\nu,c],d]}^{\text{orb}} 
  &=&\sum_{(J,J^\prime)\in\Gamma} 
     \sum_{[\sigma,e]\in\Rep(\G/\P),f\in\Rep(\P)}\:N_{[\mu,a]^+[\nu,c]}^{J
    [\sigma,e]}\,N_{b^+d}^{J^\prime f}\:\chi_{[\sigma,e]}^{\G/\P}\,\chi_f^{\P}
    \ \ . 
\qe 
When the orbifold action has fixed points some of these states may be 
resolved further, but we will not discuss this issue. The main point 
here was to outline how one can obtain branes in the background 
\eqref{eq:OrbifoldPartition}. They are labeled by elements of 
$\bigl(\Rep(\G/\P)\times\Rep(\P)\bigr)/\Gamma$ and come with an 
obvious geometric interpretation. Moreover, in \cite{Brunner:2000wx} 
one can find explicit formulas for the boundary operator product 
expansions in such boundary theories (see also \cite{Matsubara:2001hz} 
for a generalization to orbifolds with fixed points). 
\smallskip 

Let us note that 
open string theory on conformal field theory orbifolds was pioneered by 
Sagnotti and collaborators starting from \cite{Pradisi:1989xd} and 
systematized in \cite{Bianchi:1990yu,Bianchi:1991tb} (see also e.g.\ 
\cite{Bianchi:1991rd,Pradisi:1995qy,Pradisi:1995pp}). Important 
contributions were made later by Behrend et al.\ \cite{Behrend:1998fd,
Behrend:1999bn} and by Fuchs et al.\ \cite{Fuchs:1999zi,Fuchs:1999xn,
Birke:1999ik} (see also \cite{Huiszoon:1999jw,Schellekens:2000eh}).   
\smallskip

\subsection{Comparison with the new boundary states}

Our task now is to find a choice for the orbifold group $\Gamma
= \Gamma_0 \subset \Cent(\G/\P) \times \Cent(\P)$ such that the 
partition function \eqref{eq:OrbifoldPartition} coincides with 
the charge conjugated modular invariant of the 
$\G$-theory, i.e.\ 
with the expression \eqref{GPF}. In more geometric terms, the 
condition on the orbifold group is $\G = (\G/\P \times \P)/
\Gamma_0$. It will turn out that the existence of an appropriate 
group $\Gamma_0$ imposes strong constraints on the choice
of $\P$. Once these constraints have been formulated, we shall 
compare the boundary states \eqref{orbbs} with our new boundary 
states~(\ref{sol1a},\,\ref{sol1b}).
\smallskip 
 
To formulate necessary conditions for the equivalence of the 
partition function \eqref{GPF} of the $\G$-theory with one of 
the orbifold partition functions \eqref{eq:OrbifoldPartition}, 
we shall concentrate on terms that contain a factor $\chi_{[0,0]} 
\chi_0$ from the holomorphic sector,     
\eq
  Z &=& {\sum}_{a}  \ \chi^{\G/\P}_{[0,0]} \ \chi^\P_0 \ \, 
      \bar{\chi}^{\G/\P}_{[0,a]} 
       \   \bar{\chi}^\P_a  \ +\  \dots 
\ \ , \\[1mm] 
Z^{\text{orb}} &=& {\sum_{(J,J') \in \Gamma}}\  \chi^{\G/\P}_{[0,0]} 
   \ \chi^\P_0 \ \, \bar{\chi}^{\G/\P}_{J [0,0]} \
   \bar{\chi}^\P_{J'0} \ + \ \dots \ \ . 
\qe         
The summation over $a$ in the first expression is restricted such 
that $(0,a) \in \Allowed(\G/\P)$. We can now read off one important 
condition for the equivalence: all the labels $a\in \Rep(\P)$ that 
appear in the summation must be simple currents of $\A(\P)$, i.e.\ 
elements of $\Cent(\P)$. Under this condition we can set   
$$ \Gamma \:= \ \Gamma_0 \ = \ \bigl\{\bigl([0,a],a\bigr)\  \bigl| \
   (0,a) \in \Allowed(\G/\P) \bigr\} \ \subset \Cent (\G/\P) \times
   \Cent(\P) \ \ . $$ 
By projection on the first or second factor, $\Gamma_0$ can be 
identified with a subgroup of both $\Cent(\P)$ and $\Cent(\G/\P)$. If the 
identification group $\Gid$ is trivial, it follows that all sectors 
of $\A(\P)$ must be simple currents, i.e.\ $\P$ must be abelian. In 
cases with non-trivial field identification, the orbifold construction 
with $\Gamma_0$ can reproduce the partition function of the $\G$-theory
even if some of the sectors of $\P$ are not simple currents. We shall 
provide one example at the end of this section. 

A more detailed comparison of the bulk partition functions reveals 
a second necessary condition for the desired equivalence. Namely, 
one can see that the orbifold and the $\G$-theory can only agree 
if $\Gamma = \Gamma_0$ acts transitively on the sets $\Allowed_\mu
(\G/\P) := \bigl\{ (\mu,b) \in \Allowed(\G/\P)\bigr\}$. In particular,
this implies
that $\bigl|\Allowed (\G/\P)/ \Gamma_0\bigr| = |\Rep (\G)|$.   
\medskip 

We are now prepared to compare the brane spectra of the orbifold 
construction with the spectra obtained in Section 3. In the following 
analysis we assume that $\P$ is abelian which is the case for all the 
examples considered in \cite{Maldacena:2001ky,Maldacena:2001xj}. The 
orbifold construction of the background works for a slightly larger 
class of cases, but in such cases the brane spectra can be different, 
at least before resolving possible fixed points of $\Gamma_0$ 
(see below). Assuming that $\Rep(\P) = \Cent(\P)$, we want to verify 
first that both constructions provide the same number of boundary
states. This amounts to saying that 
\eqn 
  \label{eq:BSEquality} 
  \bigl(\Rep(\G/\P)\times\Rep(\P)\bigr)/\Gamma_0 
  \ \cong\ \bigl(\Rep(\G)\times\Rep(\P)\bigr)/\Gid\ \ . 
\qen 
By our assumption on $\P$, the action of $\Gamma_0$ has no fixed 
points. The same holds true automatically for the action of $\Gid$.
Therefore, our results of Section~\ref{sc:NewStates} apply and
it is easy to compute the order of the two sets 
in relation \eqref{eq:BSEquality}. For the set on the left hand side we 
find that 
$$ \Biggl| \frac{\Rep(\G/\P)\times\Rep(\P)}{\Gamma_0} \Biggr| \ = \ 
    \frac{|\Allowed(\G/\P)| \cdot |\Rep(\P)|}{|\Gamma_0|\cdot|\Gid|}
    \ = \ \frac{|\Rep(\G)|\cdot |\Rep(\P)|}{|\Gid|} \ \ . $$ 
This agrees with the number of new boundary states on the right hand side
of eq.~\eqref{eq:BSEquality}. If we drop the
assumption $\Rep(\P) = \Cent(\P)$ the action of $\Gamma_0$ 
can have fixed points so that the number of unresolved branes is 
smaller than the number of branes we obtained from our construction.
\smallskip 

To compare the open string spectra of the two sets of branes we have 
to go a step further and choose an explicit isomorphism between the 
labels. Let us propose 
\eq
  \vartheta : \bigl[[\mu,b],c\bigr] \ \mapsto \ (\mu,b-c)\ \ .
\qe
Note that $b-c \in \Rep(\P)$ makes sense for two elements $b,c \in 
\Rep(\P)$ since we assume $\Rep(\P) = \Cent(\P)$ to be an abelian
group. Furthermore, 
$\vartheta$ is well-defined because the action of $\Gamma_0$ on 
the labels $\bigl([\mu,b],c\bigr) \in \Rep(\G/\P) \times \Rep(\P)$ adds the 
same $a$ to $b$ and $c$ so that their difference $b-c$ is left 
invariant. In writing down the pair $(\mu,b-c)$ we have to pick 
a representative $(\mu,b)$ of the sector $[\mu,b]$. This is unique
up to the action of the identification group $\Gid$. But different 
representatives are mapped to the same $\Gid$-orbit in $\Rep(\G) 
\times \Rep(\P)$. Obviously, $\vartheta$ is surjective and hence, 
by our counting above, it is a bijection between the two sets of 
labels.      
\smallskip

It is now straightforward to compare the boundary partition function 
coming from our construction with those arising from the orbifold 
analysis. In the following we shall identify the elements
$\bigl([0,a],a\bigr)
\in \Gamma_0$ with $a \in \Cent(\P)$. We first calculate the boundary 
partition function from the orbifold point of view. Using the formula
\eqref{cosN} we obtain  
\eq 
  Z_{[[\mu,b_1],c_1],[[\nu,b_2],c_2]}^{\text{orb}} 
  \ =\ \sum_{\substack{a \in \Gamma_0,(\sigma,d)\in\Allowed(\G/\P)\\
   e\in\Rep(\P)}} {N_{\mu\nu}}^{\sigma}\ {N_{b_1^+\,b_2}}^{a + d}\ 
    {N_{c_1^+\,c_2}}^{a+e} 
   \ \chi_{(\sigma,d)}^{\G/\P}\:\chi_e^{(\P)}\ \ .
\qe 
In this particular example, the fusion coefficients for the $\P$-part
are well-known and parts of the sum may be carried out. A careful
calculation  leads to
\eq 
  Z_{[[\mu,b_1],c_1],[[\nu,b_2],c_2]}^{\text{orb}} 
  =\sum_{(\sigma,d)\in\Allowed(\G/\P)}{N_{\mu\nu}}^{\sigma}\:
   \chi_{(\sigma,d)}^{\G/\P}\:\chi_{d+c_2-c_1+b_1-b_2}^{(\P)}
\ \ . 
\qe 
Let us now consider the boundary partition function for the 
corresponding weights $(\mu,b_1-c_1)$ and $(\nu,b_2-c_2)$ in our 
approach. Again, a careful analysis yields
\eq
  Z_{(\mu,b_1-c_1),(\nu,b_2-c_2)} 
  &=&\sum_{\substack{(\sigma,d)\in\Allowed(\G/\P)\\ {e,f\in\Rep(\P)}}}
   {N_{\mu\nu}}^{\sigma}\  
     {N_{(b_1-c_1)^+\,(b_2-c_2)}}^{f}\ {N_{fe}}^{d}\ 
     \chi_{(\sigma,d)}^{\G/\P}\:\chi_e^{(\P)}\\[1mm] 
  &=&\sum_{(\sigma,d)\in\Allowed(\G/\P)}{N_{\mu\nu}}^{\sigma}\:
     \chi_{(\sigma,d)}^{\G/\P}\:\chi_{d+c_2-c_1+b_1-b_2}^{(\P)}\ \ . 
\qe
This agrees with the result of the orbifold construction and thus
proves the equivalence of the two approaches. 

\subsection{An instructive example} 

For our general comparison of brane spectra in the previous subsection 
we assumed that $\P$ is abelian, i.e.\ that all sectors of $\A(\P)$ are 
simple currents. This assumption was sufficient for the equivalence of 
the bulk partition functions but not necessary when the identification 
group $\Gid$ is non-trivial. In this subsection we shall present one 
such example.    
\smallskip

To begin with, let us set
$\A(\G) = \A\bigl(\asu(2)_{k_1}\oplus\asu(2)_{k_2}\bigr)$. 
This chiral algebra has several subalgebras $\A(\P)$ that we could choose
for our construction of boundary states. There are various abelian subalgebras
that we could use such as $\A(\P)=\A\bigl(\au(1)_{k_1}\bigr)$ or
$\A(\P)=\A\bigl(\au(1)_{k_1}\oplus\au(1)_{k_2}\bigr)$
etc. To make things a bit more 
interesting we shall pick a non-abelian subalgebra, namely the chiral 
algebra that is generated by the diagonally embedded subalgebra 
$\asu(2)_{k_1+k_2}$. The corresponding projection of weights is given 
by $\mc{P}(\mu,\alpha)=\mu+\alpha$. Sectors of the coset theory are 
labeled by triples $(\mu,\alpha,a)$ with $\mu\leq k_1,\alpha\leq k_2$,
$a\leq k_1+k_2$ and the branching selection rule $\mu+\alpha-a = 0 
\mod 2$. One can show that there is only one non-trivial field 
identification current $(k_1,\,k_2,\,k_1+k_2)$. It gives rise to the 
field identification
\eq
  (\mu,\alpha,a)\ \sim\ (k_1-\mu,\,k_2-\alpha,\,k_1+k_2-a)\ \ .
\qe
Since we want to avoid fixed points of the field identification we have 
to consider the situation where at least one of the levels is odd.
\smallskip

We now specialise to the case $k_1=k_2=1$ for which the coset algebra 
is the chiral algebra of the Ising model. The relevant lists of sectors 
are,  
\eq
  \Rep(\G)
   &=& \Rep\bigl(\asu(2)_1 \oplus \asu(2)_1\bigr) 
   \ = \ \bigl\{(0,0), (0,1), (1,0), (1,1) \bigr\} \\[2mm]  
  \Rep(\P) &=&  \Rep\bigl(\asu(2)_2\bigr) \ = \ \bigl\{ 0, 1, 2\bigr\} \\[2mm] 
  \Rep(\G/\P)&=&\bigl\{(0,0,0)\sim(1,1,2),\,(0,0,2)\sim(1,1,0),\,(0,1,1)\sim(1,0,1)\bigr\}\ \ .
\qe
Next, we have to decompose the charge conjugated modular invariant partition 
function for $\A\bigl(\asu(2)_1 \oplus \asu(2)_1\bigr)$ into characters
of the reduced chiral algebra. In our case this reads, 
\eqn
  \label{eq:PartitionSUSU}
  Z&=&\Bigl|\chi_{(0,0)}^{\G}\Bigr|^2+\Bigl|\chi_{(0,1)}^{\G}\Bigr|^2+
  \Bigl|\chi_{(1,0)}^{\G}\Bigr|^2+\Bigl|\chi_{(1,1)}^{\G}\Bigr|^2\\[2mm] 
  \nonumber
  &=&\Bigl|\chi_{(0,0,0)}^{\G/\P}\chi_0^{\P}+\chi_{(0,0,2)}^{\G/\P}
   \chi_2^{\P}\Bigr|^2 +2\Bigl|\chi_{(0,1,1)}^{\G/\P}\chi_1^{\P}\Bigr|^2
  +\Bigl|\chi_{(0,0,2)}^{\G/\P}\chi_0^{\P}+\chi_{(0,0,0)}^{\G/\P}
   \chi_2^{\P}\Bigr|^2
\qen
where we already took the field identification into account. 
Following the results of Section~\ref{sc:TrivialTrivial} for 
trivial gluing conditions on the reduced chiral algebra, we
find six boundary states with labels from the set  
\begin{multline*}
  \B=\bigl\{(0,0,0)\sim(1,1,2),\,(1,0,0)\sim(0,1,2),\,(0,1,0)\sim(1,0,2),\\[1mm]
   (1,1,0)\sim(0,0,2),\,(0,0,1)\sim(1,1,1),\,(1,0,1)\sim(0,1,1)\bigr\}\ \ . 
\end{multline*}
The four boundary states which are in the $\Gid$-orbit of the 
labels $(\mu,\alpha,0)$ with trivial last entry can be identified 
with the four Cardy states of the model. All of them preserve the 
full chiral algebra $\A(\G)$. For the remaining two boundary 
states we find 
\begin{align*}
Z_{(0,0,1),(0,0,1)}
  &=\chi_{(0,0,0)}^{\G/\P}\chi_0^{\P}+\chi_{(0,0,2)}^{\G/\P}\chi_2^{\P}
  +\chi_{(0,0,0)}^{\G/\P}\chi_2^{\P}+\chi_{(0,0,2)}^{\G/\P}\chi_0^{\P}
  &= \chi_{(0,0)}^{\G}+\chi_{(1,1)}^{\G}\ \\
Z_{(0,0,1),(1,0,1)}
  &=2\chi_{(0,1,1)}^{\G/\P}\chi_{1}^{\P}
    &= \chi_{(1,0)}^{\G}+\chi_{(0,1)}^{\G}\ \\
  Z_{(1,0,1),(1,0,1)}
  &=\chi_{(0,0,0)}^{\G/\P}\chi_0^{\P}+\chi_{(0,0,2)}^{\G/\P}\chi_2^{\P}+\chi_{(0,0,0)}^{\G/\P}\chi_2^{\P}+\chi_{(0,0,2)}^{\G/\P}\chi_0^{\P}
  &= \chi_{(0,0)}^{\G}+\chi_{(1,1)}^{\G}\ .
\end{align*}
In particular, these boundary conditions preserve the full chiral  
symmetry! This is rather accidental and it is related to the fact 
that $\asu(2)_1 \oplus \asu(2)_1$ possesses an outer automorphism 
which acts by exchanging the two summands. With our construction we
just recovered the two boundary states which belong to the associated 
twisted gluing condition. Note, however, that the spectra of open 
strings which stretch in between the four Cardy and the two non-Cardy 
type branes do only preserve the reduced chiral symmetry.%
\medskip%

Before we conclude this section let us observe that the partition 
function~\eqref{eq:PartitionSUSU} actually is an orbifold partition 
function obtained with the orbifold group
$$ \Gamma_0 \ = \ \bigl\{\bigl([0,0,0],0\bigr),\bigl([0,0,2],2\bigr)\bigr\}
  \ \cong \ \Integer_2 \ \ . 
$$  
In fact, the partition function of our model is recovered from the 
general expression~\eqref{eq:OrbifoldPartition} with the help of 
$Q_{([0,0,2],2)}\bigl(\bigl[[\mu,\alpha,a],b\bigr]\bigr)=(b-a)/2$
and using that the weight 
$\bigl[[0,1,1],1\bigr]$ is invariant under $\Gamma_0$. On the other hand, 
$\P = \asu(2)_2$ is not abelian since
$(1)\in\Rep\bigl(\asu(2)_2\bigr)$ is not a simple current. 
\smallskip

The orbifold group $\Gamma_0$ acts on the set $\Rep(\G/\P)\times 
\Rep(\P)$. Under this action, the nine elements of the latter 
are grouped into four orbits of length $2$ and one fixed point. 
Hence, before resolution of the fixed point one obtains five 
boundary states of the form (\ref{orbbs}). But the one brane 
$|([0,1,1],1)\rangle$ which is associated with the fixed point 
of $\Gamma_0$ can be resolved into a sum of two elementary 
branes. In this way we recover all the six branes with symmetry 
$\A(\G/\P) \oplus \A(\P)$ from the orbifold construction. Note 
that in our approach the issue of fixed point resolution did 
not arise.

\section{\label{sc:TensorProducts}Product geometries and defect lines}

Our final goal is to apply our general formalism to tensor 
products of two or more conformal field theories. Such product 
theories appear in particular whenever the background geometry 
splits into several factors. Moreover, tensor products also 
arise in the description of defect lines in 2D systems due to 
the so-called `folding trick'. We shall explain the general 
ideas behind these two types of applications in the first 
subsection and then illustrate the constructive power of our 
formulas by considering products of  WZW models. 

\subsection{Boundary states in tensor products} 

As we have just noted, there exist at least two important 
motivations for the analysis of branes in product theories. 
First of all, many string backgrounds are obtained as products 
from several factors, interesting examples for our purposes 
being $AdS_3 \times S^3 \times T^4$ or $AdS_3 \times S^3 
\times S^3 \times \Real$. Some boundary states for such theories 
can be factored accordingly so that they are simply products 
of boundary states for each of the individual factors. But 
this does not exhaust all possibilities, as one can understand
most easily by considering the simple product $S^1 \times S^1$. 
Since there are only point-like and space-filling branes on a 
circle, products of the corresponding boundary states can only 
give point-like branes, 1-dimensional branes running parallel 
to one of the two chosen circles and space filling branes with 
vanishing magnetic field. 1-dimensional branes which run 
diagonally through the 2-dimensional space and, closely related, 
space filling branes with a B-field are not factorisable. In this 
example, the factorisable branes suffice at least to generate all 
the possible RR-charges. However, this is not true for many other 
product geometries. In the case of $S^3 \times S^3$, for example, 
stable factorisable branes carry only 0-brane charge. But K-theory 
predicts the existence of additional branes with non-vanishing 
3-brane charge which can not be built up from branes on the 
factors $S^3$. Hence, there is a strong demand for additional 
boundary states. We shall show below that our ideas can be 
fruitfully applied in this context. 
\smallskip 
 
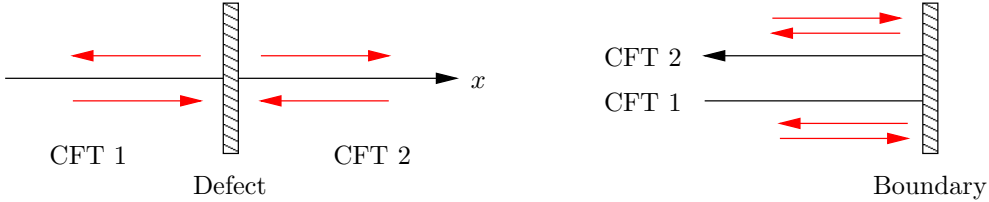
\begin{figure}
  \centerline{\input{bothlines.pstex_t}}  
  \caption{\label{fig:FoldingTrick}The folding trick relates a system on the
    real line with a defect to a tensor product theory on the half line.}
\end{figure}

There exists another -- superficially very different -- setup 
which leads to exactly the same type of problems. It arises 
by considering a one-dimensional quantum system with a defect
(see e.g.\ \cite{Oshikawa:1997dj,LeClair:1997gz,Nayak,
Saleur:1998hq,Saleur:2000gp} and \cite{McAvity:1995zd,Erdmenger:2002ex} 
for higher dimensional analogues), or, more generally, two different 
systems on the half-lines $x<0$ and $x>0$ which are in contact at 
the origin. The defect or contact at $x=0$ could be totally 
reflecting, or more interestingly it could be partially (or 
fully) transmitting. To fit such system into our general 
discussion, we apply the usual folding trick (see Figure 
\ref{fig:FoldingTrick}). After such a folding, the defect or 
contact is located at the boundary of a new system on the 
half-line. In the bulk, the new theory is simply a product 
of the two models that were initially placed to both sides 
of the contact at $x=0$.   
Factorizing boundary states for the new product theory on the 
half line correspond to totally reflecting defects or contacts.
With our new boundary states we can go further and couple the 
two systems in a non-trivial way. Since we always start with 
conformal field theories $\H_1= \H_<$ and $\H_2= \H_>$ on either 
side of $x=0$, it is natural to look for contacts that preserve 
conformal invariance. This requires to preserve the sum of the
two Virasoro algebras of the individual theories. After folding 
the system, the preserved Virasoro algebra is diagonally embedded 
into the product theory $\G = \H_1 \times \H_2$. Of course, one 
can often embed a larger chiral algebra $\P$ and then look for 
defects that preserve this extended symmetry. This is exactly 
the setup to which our general ideas apply.

\subsection{Defect lines with jumping central charge}

Our goal is now to construct examples of defect lines that join 
two conformal field theories with different central charge. Such 
situations are known to appear on the boundary of an AdS-space 
whenever there is a brane in the bulk which extends all the way to the 
boundary \cite{Karch:2000ct,Karch:2000gx,DeWolfe:2001pq,Bachas:2001vj}. 
To be specific, we will choose two WZW models based on the same 
semi-simple Lie group $\H$ but at different levels $k_i$ and hence 
with different central charges $c_i = k_i\,\dim \H / (k_i + h^\vee)$. 
The boundary states we shall discuss may also be interpreted as 
D-branes in the product geometry $\H_1 \times \H_2$ in which the two 
factors may have different volume. 
\smallskip 

In this setup, the `$\G$-theory' is provided by the charge 
conjugated modular invariant partition function for $\A(\G) = 
\A(\H_1) \oplus \A(\H_2) = \A(\ah_{k_1} \oplus 
\ah_{k_2})$. Now we are instructed to choose some chiral 
subalgebra $\A(\P)$. There are many different choices, but 
we shall  use the affine algebra $\ah_{k_1+k_2}$ which is 
embedded diagonally into $\ah_{k_1} \oplus \ah_{k_2}$. In 
other words, $\P\cong\H_D$ and $\A(\H_D) = \A(\ah_{k_1+k_2})$.  
\smallskip

We start by introducing some pieces of notation. As we have to deal
with three {\em different} affine algebras $\ah$, it is convenient 
to use different labels for the sectors of each of these algebras,  
\eq
  \ah_{k_1}: \:\mu,\nu,\ldots \ \in \ \Rep(\H_1) \ \ ,  \qquad
  \ah_{k_2}: \:\alpha,\beta,\ldots \ \in \ \Rep(\H_2) \ \  , 
   \\[2mm] 
  \ah_{k_1+k_2}: \:a,b,\ldots \ \in \ \Rep(\H_D)\ \ . 
  \quad \quad \quad \quad \quad \quad
\qe
In the case under consideration, the projection is given by $\mc{P}
(\mu,\alpha)=\mu+\alpha$ and hence the branching selection rule 
\eqref{eq:BranchingSelection} reduces to $\mu+\alpha-a\in \mc{Q}$ where
$\mc{Q}$ denotes the root lattice of $\h$. Consequently, the coset 
fields are labeled by triples
\eq
  \bigl((\mu,\alpha),a\bigr)
  \ \in\ \Rep(\H_1\times\H_2\times\H_D)
  \quad\text{ with }\quad\mu+\alpha-a\in \mc{Q}
\qe
which give rise to a set $\Allowed(\H_1\times\H_2/\H_D)$. Next we have 
to describe the field identifications.
Let $\Cent(\H_i)$ be the centers of the $\H_i$-theories which
-- under our assumptions -- are all ismorphic.
The common center is given by the diagonal subset
\eq
  \bigl((J,J),J\bigr)\in\Cent(\H_1 \times \H_2)\times\Cent(\H_D)\ 
\qe
and leads us to the identification rules 
\eq
  \bigl((J\mu,J\alpha),Ja\bigr)\sim\bigl((\mu,\alpha),a\bigr)\ \ .
\qe
Note in particular that no additional field identifications occur
even in the case where the levels coincide, $k_1=k_2=k$, and the
two types of fields $\mu,\alpha$ take values in the same set.
\smallskip

After this preparation we can address the issue of field identification 
fixed points and spell out conditions for their absence. For the moment, 
let us focus on one of the factors and  denote it by $\H$.  Every outer 
automorphism $J\in\Cent(\H)$ is 
associated with a unique permutation $\pi_J$ of affine fundamental 
weights. Denoting affine weights by square brackets, this action may 
be written as
\eq
  J\:[\:\lambda_0\:,\:\ldots\:,\:\lambda_r\:]
  \:=\:[\:\lambda_{\pi_J(0)}\:,\:\ldots\:,\:\lambda_{\pi_J(r)}\:]\ \ .
\qe
Thus the existence of a field identification fixed point is equivalent
to finding an affine weight such that $\lambda_i=\lambda_{\pi_J(i)}$ 
for all $i=0,\ldots,r$ and at least one non-trivial $J\in\Cent(\H)$. 
The condition for the existence of such weights have been studied 
and the results for all simple Lie algebras are summarized in Table
\ref{tab:AllowedLevels}. Note that the exceptional groups $E_8$, $F_4$
and $G_2$ have trivial centers and thus no field identification or
selection rules.

To illustrate the rules summarized in Table \ref{tab:AllowedLevels}, 
let us derive them for the special case of $A_2$. The group $\Cent\bigl(
A_2^{(1)}\bigr) \cong\Integer_3$ is generated by the shift 
$$ J\:[\:\lambda_0,\lambda_1, \lambda_2\:]\  = \ 
 [\:\lambda_2,\lambda_0,\lambda_1\:] \ \ . $$ 
In terms of non-affine weights, this action reads $J\:(\:\lambda_1,
\lambda_2\:)\:=\:(\:k-\lambda_1-\lambda_2,\lambda_1\:)$. Hence, a 
fixed point would have to satisfy $\lambda_1=\lambda_2$ and $\lambda_1
=k-\lambda_1-\lambda_2$, i.e.\ it should be given by $(k/3,k/3)$. 
Obviously this is not an element of the weight lattice unless the
level~$k$ is a multiple of three. 
\smallskip

Except from the B-series, we can always find levels for which 
the action of the center $\Cent(\H)$ on the 
weights has no fixed points. In the context of our construction, 
we have three different sets of labels on which this groups acts 
at the same time and it is sufficient if at least one of the 
values $k_1,k_2$ or $k_1+k_2$ avoids the values specified in 
Table~\ref{tab:AllowedLevels}. In the following we shall assume
that this condition is satisfied. Otherwise one would have to 
resolve the fixed points according to~\cite{Fuchs:1996tq}
which leads to technical difficulties but no conceptually 
new insights.
\smallskip

The rest is now straightforward. Note that the modular S 
matrix of the `numerator theory' factorizes according to 
\eq
  S^{\H_1 \times \H_2}_{(\mu,\alpha)(\nu,\beta)}\ =
    \ S^{\H_1}_{\mu\nu}\ S^{\H_2}_{\alpha\beta}\ \ .
\qe
In this situation, the Verlinde formula~\eqref{eq:VerlindeFormula} 
implies that the same holds for the fusion coefficients
\eq
  N_{(\mu,\alpha)(\nu,\beta)}^{(\:\rho,\gamma)}
  \ =\ {N_{\mu\nu}}^{\rho}\:{N_{\alpha\beta}}^{\gamma}\ \ .
\qe
Our boundary states are now labeled by $\Gid$-orbits of
triples $\bigl((\rho,\gamma),r\bigr)$. When we finally insert these 
expressions into the formula~\eqref{eq:CosetOpenPartition}, we can 
read off their boundary partition function, 
\eq
  Z(q)\ =\ \sum_{((\nu,\beta),b),c,d}\:\Bigl[{N_{\rho_1^+\rho_2}}^{\nu}
  {N_{\gamma_1^+\gamma_2}}^{\beta}{N_{r_1^+r_2}}^{d}{N_{dc}}^{b}\Bigr]
  \:\chi_{((\nu,\beta),b)}^{\H_1\times\H_2/\H_D}(q)\:\chi_{c}^{\H_D}(q)\ \ .
\qe
When reinterpreted in terms of defects, these formulas provide us with 
a large set of possible junctions between two conformal field theories. 
Note that these have different central charge if $k_1 \neq k_2$.  
\begin{table}
\centerline{\begin{tabular}{c|cccccc}
  Algebra & $A_{n}^{(1)}$ & $B_{n}^{(1)}$ & $C_{n}^{(1)}$ & $D_{n}^{(1)}$ & $E_{6}^{(1)}$ & $E_{7}^{(1)}$ \\\hline
  FPs for $k$ in & $\bigcup\limits_{1\neq s|(n+1)}\!\!\!\!\!\!s\Natural_0$ & $\Natural_0$ & $2\Natural_0$ & $2\Natural_0$ & $3\Natural_0$ & $2\Natural_0$
  \end{tabular}}
  \caption{\label{tab:AllowedLevels}Existence of fixed points under simple current actions.}
\end{table}

\section{Conclusions and Outlook}

In this work we proposed a new algebraic construction of symmetry 
breaking boundary states of some given bulk conformal field theory
-- the $\G$-theory -- with chiral algebra $\A(\G)$. According to our 
prescription, one starts by choosing some rational subalgebra $\A(\P)$ 
of the full chiral algebra $\A(\G)$ together with a gluing automorphism 
$\Omega^\P$. Using the solution of the boundary problem for an auxiliary 
$\P$-theory with the gluing automorphism $\Omega^\P$, we were able to 
build new boundary states for the $\G$-theory (see formula \eqref{sol2}). 
For the simplest choice $\Omega^\P= \id$, the auxiliary $\P$-theory
is solved by Cardy's solution. In this way one obtains at least 
$|\Rep(\G)| \cdot |\Rep(\P)|/|\Gid|$ boundary states of the $\G$-theory 
for each admissible $\P$ (see eqs.\ (\ref{sol1a},\,\ref{sol1b})). If all 
sectors of $\A(\P)$ are simple currents, i.e.\ if $\A(\P)$ is abelian, 
then our `algebraic' boundary states coincide with the boundary states 
\eqref{orbbs} which can be obtained from an orbifold construction of 
the type suggested in \cite{Maldacena:2001ky,Maldacena:2001xj}. But 
our formulas do not require~$\P$ to be abelian and hence they provide 
a true generalization of the orbifold ideas. 
\smallskip

We have also discussed how the new boundary states can be applied 
to obtain non-factorizing (`diagonal') branes in product geometries 
or, equivalently, to the construction of non-trivial defect lines in 
2D conformal field theory. We presented examples in which two CFTs 
with different central charge are joined along the defect line. Such 
phenomena are known to appear in the AdS/CFT correspondence whenever
branes extend to the boundary of the AdS-space \cite{Karch:2000ct,
Karch:2000gx,DeWolfe:2001pq,Bachas:2001vj}.
The jump of the central charge along the defect depends 
on the charges of the brane. 
\smallskip 

Following \cite{DeWolfe:2001pq,Bachas:2001vj}, 
it would be interesting to compute the Casimir energy between two defects. 
In a stationary system the defects arrange the excitation modes 
such that the energy density between the defects is lower than 
in the outside region resulting in an attractive force between 
the defects. This is well-known for the electro-magnetic field 
between two conducting plates. In~\cite{Bachas:2001vj}, the 
Casimir energy was calculated for a free boson system with 
defects which join regions with different compactification 
radii. 
\medskip 

Let us stress that the number of new boundary states can be enlarged 
by a simple iteration of our construction. In addition to the chiral 
symmetry $\A(\P)$ we have decided to preserve the coset algebra 
$\A(\G/\P)$ so as to render the boundary problem rational. But it 
would be possible to reduce the symmetry even further by choosing 
a rational chiral subalgebra $\A(\P') \embin \A(\G/\P)$ which 
should then be preserved together with the chiral algebra $\A(\G/
\P/\P')$ of the `double coset'.
\smallskip

It is well known that in many backgrounds, e.g.\ $\SU(N\geq4)$ or 
$S^3 \times S^3$, the standard constructions of boundary conformal 
field theory do not suffice to generate the whole lattice of 
RR-charges. Having obtained a very large class of new symmetry 
breaking boundary states, the situation is likely to improve 
drastically. It would be interesting to study more examples 
and to understand which charges are carried by our new branes.  
\smallskip

Finally, let us also stress once more that we have only provided 
a list of new boundary states. These contain information about how
closed strings couple to the associated branes and about the spectrum 
of open string modes. The operator product expansions of open string 
vertex operators (or boundary fields) contain additional data. They 
can be obtained as solutions of the sewing constraints which have 
been worked out by several authors \cite{Lewellen:1992tb,Pradisi:1996yd,
Runkel:1998pm,Behrend:1999bn,Felder:1999mq}. These constraints were 
solved for a series of orbifold models in \cite{Runkel:1999dz} and 
then more systematically for simple current orbifolds in 
\cite{Matsubara:2001hz}. Using our discussion from Section 4, the 
formalism of \cite{Matsubara:2001hz} can be applied to our present 
context whenever $\P$ is abelian and it provides the desired boundary 
operator product expansions. For more general choice of $\P$, 
however, the problem remains to be solved.

\subsection*{Acknowledgements}

We wish to thank C.~Bachas, S.~Fredenhagen, A.~Recknagel,
I.~Runkel, Ch. Schweigert and J.B.~Zuber for interesting discussions.  
The work of TQ was supported by the Studienstiftung des deutschen 
Volkes. VS and TQ acknowledge the warm hospitality of the ENS Paris
and the LPTHE Jussieu where this work was initiated. The stay of TQ
was supported by the European Superstring Theory Network.

\newpage
\noindent{\em Note added in proof:}
After this paper was completed we noticed that
our constructions provide branes which are localized along the sets
\begin{equation*}
  \mc{C}_\rho^{\G,\Omega}\bigl(\mc{C}_r^{\P,\Omega^{\P}}\bigr)^{-1}
  \subset\G\ \ ,
\end{equation*}
  where $\mc{C}_\rho^{\G,\Omega}$ and $\mc{C}_r^{\P,\Omega^{\P}}$
  are twisted conjugacy classes. Details will appear elsewhere
  \cite{TQVS:Unpublished}.
  

\providecommand{\href}[2]{#2}\begingroup\raggedright\endgroup

\end{document}

%% file: bothlines.pstex_t
\begin{picture}(0,0)%
\epsfig{file=bothlines.pstex}%
\end{picture}%
\setlength{\unitlength}{4144sp}%
\begingroup\makeatletter\ifx\SetFigFont\undefined%
\gdef\SetFigFont#1#2#3#4#5{%
  \reset@font\fontsize{#1}{#2pt}%
  \fontfamily{#3}\fontseries{#4}\fontshape{#5}%
  \selectfont}%
\fi\endgroup%
\begin{picture}(5847,1177)(-11,-686)
\put(5536,-646){\makebox(0,0)[b]{\smash{\SetFigFont{10}{12.0}{\familydefault}{\mddefault}{\updefault}Boundary}}}
\put(4051,-151){\makebox(0,0)[rb]{\smash{\SetFigFont{10}{12.0}{\familydefault}{\mddefault}{\updefault}CFT 1}}}
\put(4051,119){\makebox(0,0)[rb]{\smash{\SetFigFont{10}{12.0}{\familydefault}{\mddefault}{\updefault}CFT 2}}}
\put(1351,-646){\makebox(0,0)[b]{\smash{\SetFigFont{10}{12.0}{\familydefault}{\mddefault}{\updefault}Defect}}}
\put(2791,-16){\makebox(0,0)[lb]{\smash{\SetFigFont{10}{12.0}{\familydefault}{\mddefault}{\updefault}$x$}}}
\put(2431,-466){\makebox(0,0)[rb]{\smash{\SetFigFont{10}{12.0}{\familydefault}{\mddefault}{\updefault}CFT 2}}}
\put(271,-466){\makebox(0,0)[lb]{\smash{\SetFigFont{10}{12.0}{\familydefault}{\mddefault}{\updefault}CFT 1}}}
\end{picture}